\documentclass[aps,prl,twocolumn,showpacs]{revtex4}
\usepackage{amsmath}
\usepackage{graphicx}

\begin{document}

\title{Various spin-polarization states beyond the maximum-density droplet:\\
       a quantum {M}onte {C}arlo study} 

\author{S.~Siljam\"aki}
\author{A.~Harju}
\author{R.~M.~Nieminen}
\affiliation{Laboratory of Physics, Helsinki University of Technology, 
             P.O. Box 1100, FIN-02015 HUT, Finland}

\author{V.~A.~Sverdlov}
\affiliation{Department of Physics and Astronomy, 
             State University of New York, Stony Brook, NY 11794-3800, USA}

\author{P.~Hyv\"onen}
\affiliation{Department of Physical Sciences, Theoretical Physics, 
             University of Oulu, P.O.~Box~3000, FIN-90401 Oulu, Finland}

\date{\today}

\begin{abstract}
Using variational quantum Monte Carlo method, the effect of Landau-level 
mixing on the lowest-energy--state diagram of small quantum dots is studied 
in the magnetic field range where the density of magnetic flux quanta just 
exceeds the density of electrons.
An accurate analytical many-body wave function is constructed for various
angular momentum and spin states in the lowest Landau level, 
and Landau-level mixing is then introduced using a Jastrow factor. 
The effect of higher Landau levels is shown to be significant; 
the transition lines are shifted considerably 
towards higher values of magnetic field 
and certain lowest-energy states vanish altogether.
\end{abstract}



\maketitle

Quantum dots (QD) are systems containing a number of charge carriers 
in a nanoscale volume.
A two-dimensional semiconductor QD \cite{reed1988PRL535} 
can be constructed \textrm{e.g.}\ by superimposing a confining potential 
on two-dimensional electron gas, which can be fabricated into the 
inversion layer between two different semiconductor materials.
In the inversion layer, the confinement perpendicular to the  
interface can be made so strong that only the lowest-energy 
eigenstate in that direction has a nonnegligible probability of being occupied. 
This makes the system genuinely two-dimensional, and leads to some physical 
effects that are absent in higher dimensions: in thermodynamical limit
the integer and fractional quantum Hall effects \cite{klitzing1980PRL494}
are perhaps the most famous examples.

Experimentally,
QDs are observed to have a series of different 
ground states as the magnetic field is increased \cite{ashoori1993PRL613}.
While many properties of the states can be measured,
understanding their nature is still one of major 
theoretical goals in the field. 
In a large range of moderately high magnetic field the very stable and
fully spin polarized maximum-density--droplet state (MDD) 
remains the ground state. 
This many-particle state has an extremely accurate 
analytical description \cite{laughlin1983PRL1395}, 
and is here used as the starting point  to 
study phase diagrams of possible lowest-energy states
in stronger fields.

The variational quantum Monte Carlo (VMC) \cite{foulkes2001RMP33} 
method is used to study the QD system. 
In this method a ${\mathbf{r}}$ trial many-body wave function 
$\Psi(\{{\mathbf{r}}_i\},\{\alpha_i\})$
with desired properties and 
with free variational parameters $\alpha_i$ is first constructed, and then the
parameters are optimized
to converge towards the exact wave function $\Psi_0$. 
Using the optimized wave function, the expectation value of an observable ${\mathcal{A}}$
can be evaluated as the average of the corresponding local quantity 
$\Psi^{-1} {\mathcal{A}} \Psi$, \textrm{e.g.}, for the Hamiltonian operator ${\mathcal{H}}$:
\begin{equation}
  E_{\Psi} = \lim_{M \rightarrow \infty}
  \frac{1}{M} \sum_{i=1}^{M} \frac{{\mathcal{H}} \Psi({\mathbf{R}}_i)}{\Psi({\mathbf{R}}_i)}
  =
  \left< \Psi \left| {\mathcal{H}} \right| \Psi \right>  \,,
\end{equation}  
where the $N$-particle--coordinate configurations ${\mathbf{R}}_i$ are 
distributed as $|\Psi|^2$ and generated using the Metropolis 
algorithm.

The variational principle guarantees that the total energy
given by the VMC method, using any trial wave function, is always an upper
bound for the true total energy of the quantum state in question.
The variance of the local energy $\Psi^{-1} {\mathcal{H}} \Psi$ 
diminishes as the trial wave function approaches an eigenstate of 
the Hamiltonian, and as a result it can be used
not only as a measure of the statistical error in $E_{\Psi}$,
but also as a measure of the difference between calculated and true energies
$E_{\Psi} -E_{\Psi_0}$.

The variational parameters in the trial wave function are optimized by
minimizing the
total energy.
The minimization process itself was done using the stochastic gradient
method  \cite{harju1997PRL1173} with analytical expressions for derivatives 
 \cite{lin2000JCP2650}.
The method has proven to be fast in finding a minimum of a function
whose values are not exact.

The QD in this study is an $N$-electron system 
on a two-dimensional plane, in rotationally symmetric and parabolic
potential $V(r) = {\textstyle{\frac{1}{2}}} m^{\ast} \omega_{0}^2 r^2$, and in 
perpendicular magnetic flux density 
${\mathbf{B}} = B {\mathbf{u}}_{z}$.
An effective mass $m^{\ast}$ is used to describe the effects of the
underlying crystal structure, 
and $\omega_{0}$ determines the strength of the in-plane confinement.
In the symmetric gauge, 
${\mathbf{A}}={\textstyle{\frac{1}{2}}} {\mathbf{B}} \times {\mathbf{r}}$,
the effect of the  vector potential on the Hamiltonian
is to enhance the confinement strength, 
$\omega_{0}^2 \rightarrow \omega_{0}^2 + (\omega_{\mathrm{c}}/2)^2 = \omega^2$, 
where $\omega_{\mathrm{c}} = e B/m^{\ast}$ is the effective cyclotron frequency, 
and to introduce angular-momentum and spin-dependent terms. 
Switching to the effective harmonic oscillator units 
($m^{\ast} = \hbar = \omega = e= 1$),
the total Hamiltonian operator can be written as
\begin{equation} 
  {\mathcal{H}}  = 
  {{\textstyle{\frac{1}{2}}}}  \sum_{i=1}^N \left[ 
    - \nabla^2_i 
    + r^2_i 
    + \omega_{\mathrm{c}} (l_{z,i} 
    + \gamma^{\ast} s_{z,i} )
  \right]  
  + \sum_{i<j} \frac{C}{r_{ij}}
\, .
\label{eq:hamiltonian}
\end{equation}
Here $l_{z,i}$ and $s_{z,i}$ are the $z$-components of 
angular-momentum and spin operators for the $i$th particle, $\gamma^{\ast}$ is defined as 
$\gamma^{\ast} = g^{\ast}m^{\ast}/m_0$, where $g^{\ast}$ is the effective Land\'e g-factor,
and $C$ is 
a dimensionless interaction strength,
$C^2 = \text{Ha}/\hbar\omega \cdot m^{\ast}/(\epsilon_{\mathrm{r}}^2 m_0)$.
In these units $\omega_{\mathrm{c}} = B < 2$.

Single-particle states of the noninteracting part of the Hamiltonian (\ref{eq:hamiltonian}) can 
be solved analytically for arbitrary magnetic flux density $B$ \cite{fock1930ZP126}.
As $B$ is increased the states separate to Landau levels
with level spacing asymptotically proportional to $B$, and intra-level energy spacing 
of states asymptotically proportional to $B^{-1}$. 
In the limit of infinite magnetic flux density only the lowest Landau level (LLL) 
remains occupied. The single-particle wave functions $\varphi_{m}$,
that the LLL is composed of, have a particularly simple form:
\begin{equation}
  \label{eq:spstates}
  \varphi_{m}(z) = z^{|m|} \exp ( {-\textstyle{\frac{1}{2}}|z|^2} ) \,,
\end{equation}
with energies (in units of $\hbar \omega$)
\begin{equation}
  \label{eq:spenes}
  E_{m} = 1 + (1-{\textstyle{\frac{1}{2}}} \omega_{\mathrm{c}})\,  m\, .
\end{equation}
In Eq.\ (\ref{eq:spstates}), $z$ is a complex coordinate in the plane of the electrons: 
$z = x + {\mathrm{i}} y $, and
$m = 0,1,\dots $ is $l_{z}/\hbar$.

It is a common approximation to truncate the full Hilbert space
to the lowest Landau level only, and ignore the effects of higher
Landau levels 
altogether.
For example,  exact
diagonalization calculations are done within the LLL
approximation for $N \gtrsim 4$.
The approximation improves with increasing $B$,
since higher Landau levels move farther off in energy, but
as will be shown, Landau-level mixing (LLM) can qualitatively alter the behavior
of the system even in such regions of $B$, where the LLL approximation
is frequently used.

In this study, the trial wave functions $\Psi$ are of the 
Jastrow-Slater form. The construction of $\Psi$ begins by
creating an LLL-many-particle
wave function $\Psi_{\text{LLL}}$, which determines, \textrm{e.g.}, the angular
momentum and the spin of the state.  The LLM is then
introduced using two-body correlation functions of Jastrow
type \cite{jastrow1955PR1479}.
To construct the LLL part of the wave function, the unnormalized maximum-density 
droplet state $\Psi_{\text{MDD}}$ is used as a starting point:%
\begin{equation}
  \Psi_{\text{MDD}} = \prod_{i<j} ( z_j - z_i ) 
  \,  \exp ( -\tfrac{1}{2} 
               \sum_{i=1}^N | z_i | ^2) \, .
\end{equation}
In the thermodynamical limit ($\omega_{0} \rightarrow 0$) this configuration
corresponds to the very stable quantum-Hall state at filling factor 
$\nu=N/N_{\Phi}=1$ \cite{laughlin1983PRL1395}.
For QDs in the LLL approximation, $\Psi_{\text{MDD}}$ is the lowest-energy state in 
a large region around $\nu=L_{\text{MDD}}/L=1$. 
The state is composed of consecutive single-particle states of the 
form of Eq.\ (\ref{eq:spstates}) with $m=0,\dots,N-1$, and
it has the total angular momentum $L_{\text{MDD}} = {\textstyle{\frac{1}{2}}} N(N-1)$.

In this study only post-MDD states are considered, that is, 
states that have nonnegative additional angular momentum 
$\Delta M = L - L_{\text{MDD}}$ with respect to the MDD state.
Since the parabolic form of the external confinement allows
one to perform the elimination of the center-of-mass motion,
the spin-up and spin-down degrees of freedom are not independent. 
Therefore it is assumed that it is favorable for an electron to put 
zeros of the wave function on the positions of other spin type 
electrons also. This is equivalent to the requirement that
any post-MDD state should contain the $\Psi_{\text{MDD}}$ as a factor:
\begin{equation}
  \label{eq:mddpoly}
  \Psi_{\text{LLL}} = \Psi_{\text{MDD}} \,  P_{\Delta M} \,,
\end{equation}
where $P_{\Delta M}$ is a polynomial of proper symmetry and degree $\Delta M$.
Under this assumption, constructing the LLL trial wave function 
is a matter of choosing the suitable polynomial $P_{\Delta M}$. 
It must have the correct symmetry, and its degree $\Delta M$ gives 
the additional angular momentum with respect to the MDD state.  

\begin{figure*}[thbp]
\begin{center}
\noindent%
\includegraphics[width=6.5cm]{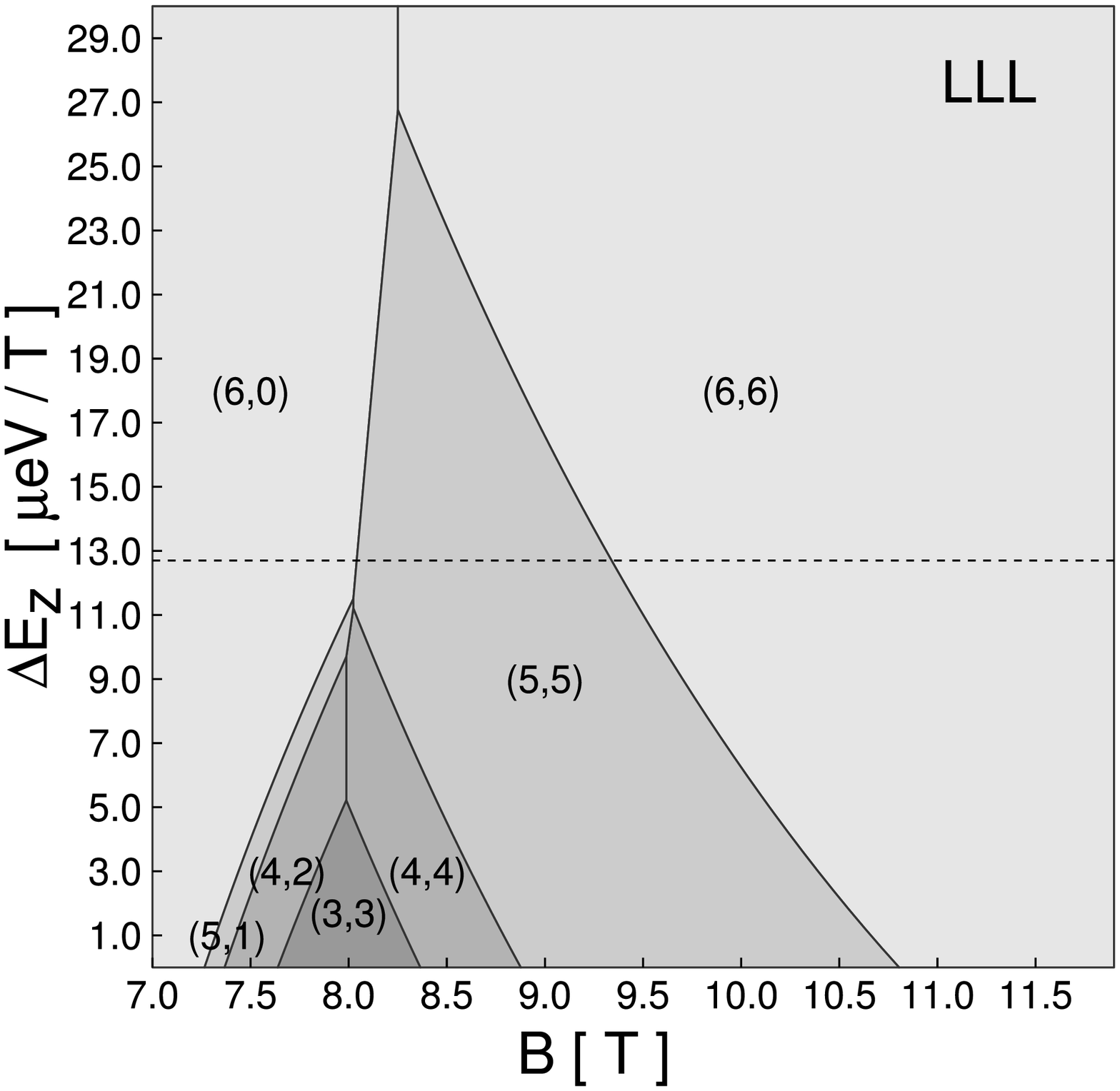}\qquad
\includegraphics[width=6.5cm]{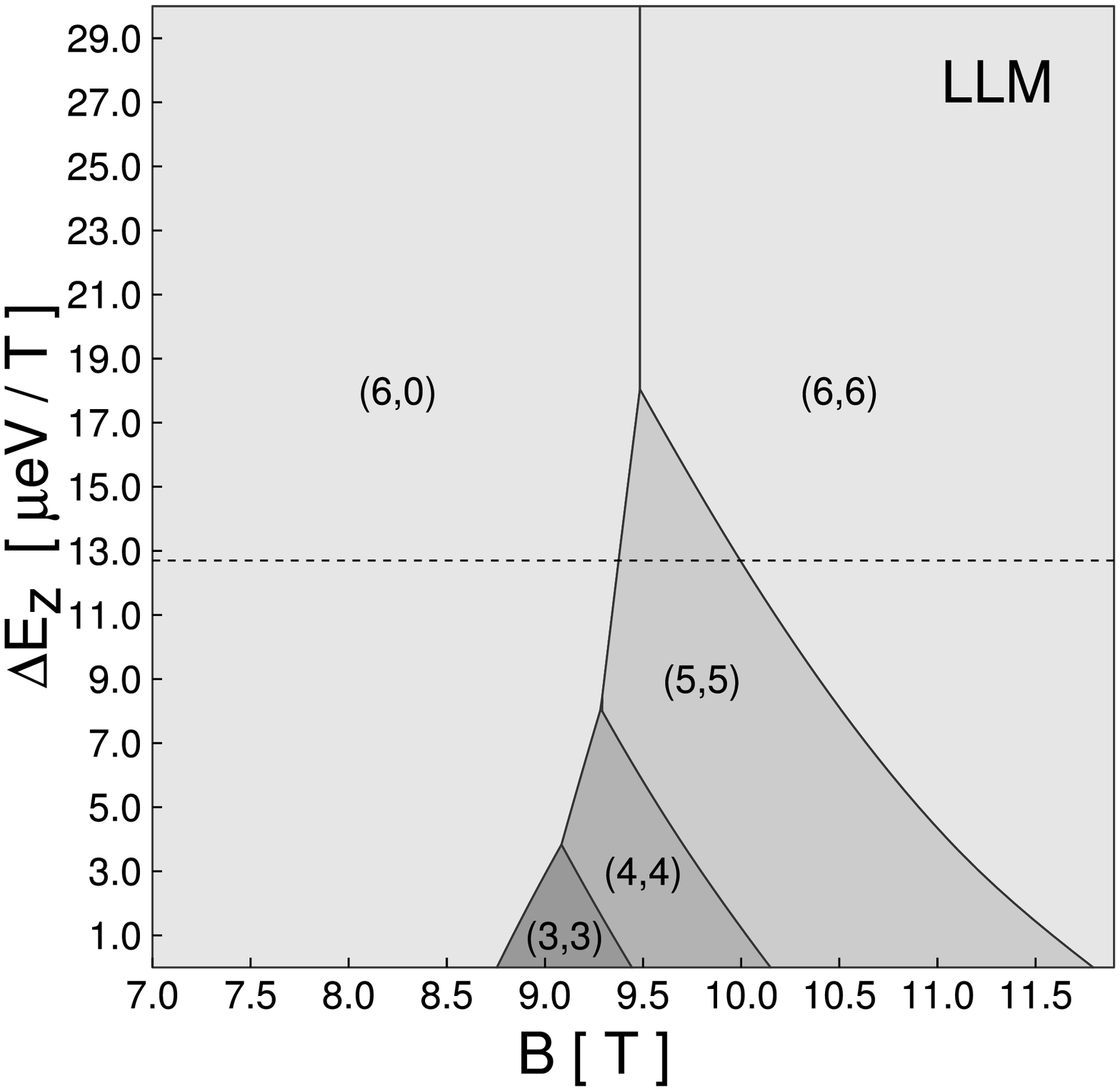}
\caption{Phase diagrams for six electrons in the lowest Landau level approximation
         (left panel) and with Landau-level mixing (right panel). 
         The labeling of the states is: $N_{\uparrow}, \Delta M$, where 
         $N_{\uparrow}$ is the number of spin-up electrons and $\Delta M$
         is the additional angular momentum ($L_{\text{MDD}}=15$ for six electrons).
         The vertical axis is the strength of the Zeeman coupling per spin,
          $\Delta {E_z} = |\frac{1}{2} \mu_{\mathrm{B}} g^{\ast}|$,
         the value of which in GaAs ($12.7\, \mathrm{\mu eV/T}$) is marked by dashed lines in the figures.
         Other parameters were: 
         $m^{\ast}/m_0 = 0.067$, $\hbar \omega_{0} = 5\, \mathrm{meV}$ and $\epsilon_{\mathrm{r}} = 12.4$.
         The relative interaction strength $C$ varies from 1.23 ($B=7\, \mathrm{T}$) to 
         1.01 ($B=12\, \mathrm{T}$).
}
  \label{fig:sixphases}
\end{center}
\end{figure*}

The polynomial $P_{\Delta M}$ is constructed as follows: 
One starts from the product
\begin{equation}
  P_{\Delta M}^{0} = 
  \prod_{i=1}^{\Delta M} \tilde{z}_i \,,
\end{equation}
where the coordinate transformation 
$\tilde{z}_i = z_i- \textstyle{{\frac{1}{N}\sum_j{z_j}}}$
has been applied to remove the center-of-mass motion. 
This fixes the additional angular momentum $\Delta M$ of the trial state. 
Correct symmetry is now built in using the Young symmetrization 
operator ${\mathcal{Y}}_\chi$:%
\begin{equation}
  \Psi_{\text{LLL}} = {\mathcal{Y}}_\chi ( \Psi_{\text{MDD}} \, P_{\Delta M}^{0} ) 
         = \Psi_{\text{MDD}} \, {\mathcal{Y}}_\chi' \, P_{\Delta M}^{0} \,,
\end{equation}
where ${\mathcal{Y}}_\chi' P_{\Delta M}^{0}$ is $P_{\Delta M}$ above.  
The parameter $\chi$ is the number of inverted spins with respect 
to a fully spin-polarized state, and it determines the shape of the 
Young tableau corresponding to ${\mathcal{Y}}_\chi$.  Because an antisymmetric 
factor $\Psi_{\text{MDD}}$ was taken out of ${\mathcal{Y}}_\chi$, the directions of 
symmetrization and antisymmetrization are interchanged in ${\mathcal{Y}}_\chi'$.
By fixing the spin and angular momentum through parameters $\Delta M$ and $\chi$,
the LLL part of the trial wave function is determined completely, \textrm{i.e.}, it has no
variational parameters.
This family of trial wave functions covers most spin configurations in the region 
$\Delta M = 0, \ldots N$. 
For example,  for seven electrons there are only three states 
(out of 23 nondegenerate states) that are 
unavailable: $S=1/2$ at $L=27$, and $S=3/2$ and $S=1/2$ at $L=28$.
These inaccessible configurations are not expected to be lowest-energy states in any case.
For states with $2S = | N -2\Delta M |$, the LLL construction above results
in wave functions that are very similar to skyrmion states \cite{macdonald1996PRL2153}.

After the lengthy construction of the LLL part, 
the LLM part is simply multiplied into the
trial wave function as a  two-body correlation function of Jastrow type
\begin{gather}
  J(\{r_{ij}\}_{i<j}) = 
  \prod_{i<j} \exp \left( \frac{\beta_{ij} r_{ij}}{  1+ \alpha_{ij} r_{ij} }
  \right) \,. \\
\intertext{The total wave function is then}
  \Psi = \Psi_{\text{LLL}} \, J \,.
\label{eq:jastr}
\end{gather}
This form of the correlation factor leaves the 
spin and angular momentum properties of $\Psi$ intact.
The Jastrow factor has two adjustable parameters: $\alpha_{ij} = \alpha_{\uparrow \uparrow}$
for parallel spins and $\alpha_{ij} = \alpha_{\uparrow \downarrow}$ for
antiparallel spins. 
For each trial state, these parameters are functions of $B$, and have to be optimized separately 
for each magnetic-field value. 
Constants $\beta_{ij}$ are determined by cusp conditions \cite{kato1957CPAM151}.
The form of Eq.\ (\ref{eq:jastr}) has proven very efficient in capturing
the Landau-level mixing;  in small systems as much 98\% of the correlation
energy can be recovered \cite{harju1998EPL407}.

\begin{figure*}[thbp]
\begin{center}
\noindent%
\includegraphics[width=6.5cm]{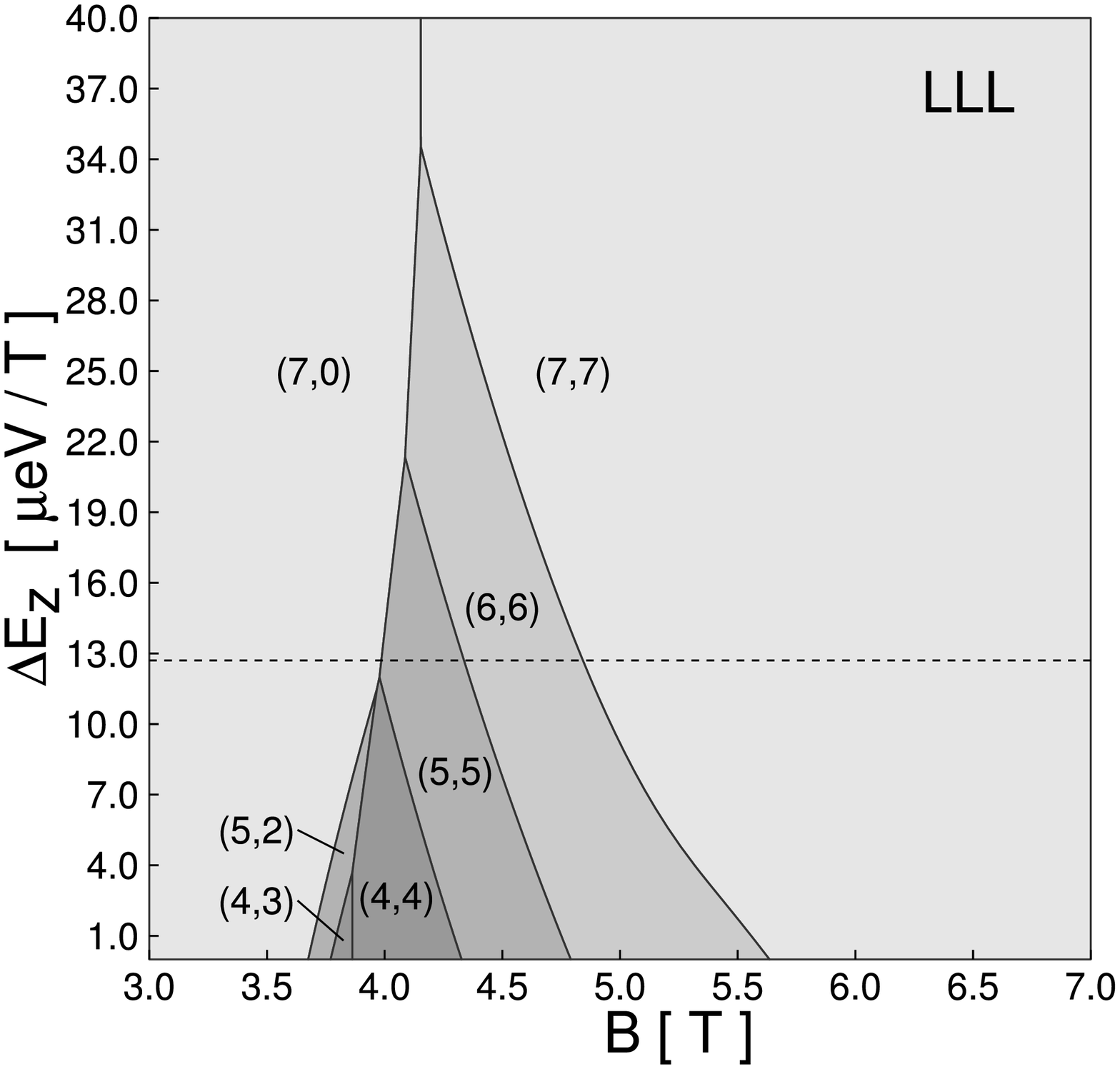}\qquad
\includegraphics[width=6.5cm]{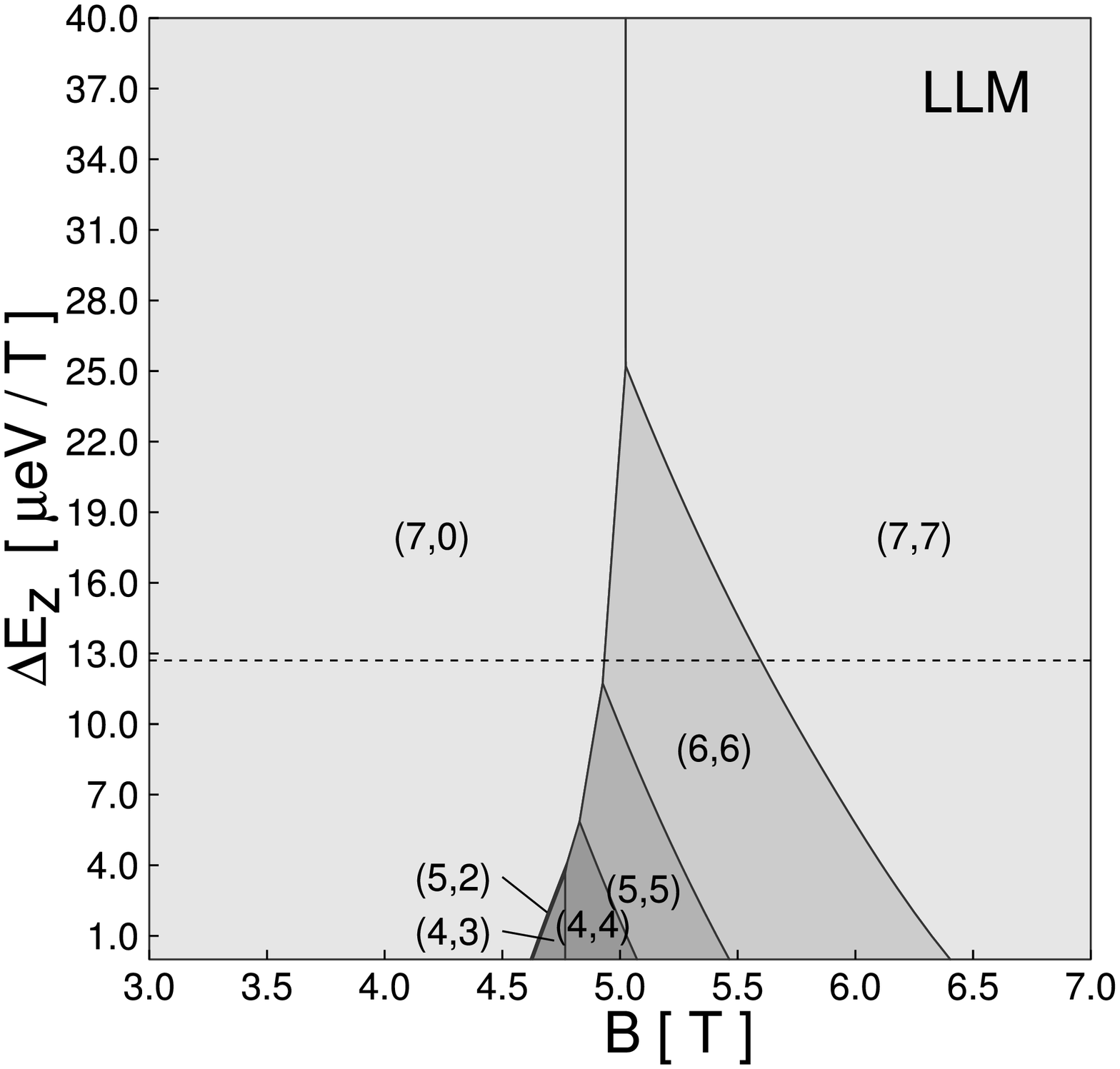}

\caption{Phase diagrams for seven electrons. 
         In this case $L_{\text{MDD}} =21$, and the value $\epsilon_{\mathrm{r}} = 13.0$ was used
         to compare with an exact diagonalization calculation. 
         The relative interaction strength $C$ varies from 1.38 ($B=3\, \mathrm{T}$) to 
         1.17 ($B=7.0\, \mathrm{T}$).
        }
  \label{fig:sevenphases}
\end{center}
\end{figure*}

Results for two systems, $N=6$ and $N=7$,  are presented in this article.
In order to test the accuracy of the LLL construction an exact diagonalization
calculation \cite{siljamaki2000PHYSB1776}
was performed for seven electrons at $B=5\,\mathrm{T}$.
This showed that for all possible LLL lowest-energy states the error 
in total energy is very small, at most 0.065\%.  
In the case of the LLM wave function, no such diagonalizations are currently available.

Fig.~\ref{fig:sixphases} shows results for a six-electron QD, 
and corresponding data for seven electrons are shown 
in Fig.~\ref{fig:sevenphases}.  Both systems use GaAs parameters: 
$m_{\text{rel}} = 0.067$, $g^{\ast} = -0.44$,
and 
$\epsilon_{\mathrm{r}} = 13.0$ for $N=7$ as in \cite{siljamaki2000PHYSB1776}%
, but
$12.4$ for $N=6$ as in \cite{saarikoski2001preprint}.%

The figures show magnetic flux density \textrm{vs.}\ Zeeman coupling strength $\Delta {E_z} = |\tfrac{1}{2}\mu_{\mathrm{B}}g^{\ast}|$.
Generally, with increasing Zeeman coupling strength the degree of
spin polarization increases, and above some critical value $\Delta E^{\ast}_z$ 
the system remains spin-polarized at all values of $B$ in the post-MDD region.
In the LLL approximation, this happens at $\Delta E^{\ast}_z=26.5\,\mathrm{\mu eV/T}$
for $N=6$, and at $34.5\,\mathrm{\mu eV/T}$ for $N=7$.
Lowering the Zeeman coupling strength 
allows other total-spin values, and with $\Delta {E_z} = 0$ the system 
goes through all possible spin configurations.  The lowest-energy states 
are the skyrmionic states as discussed above.

The effect of the Landau-level mixing can be seen on the right-hand panels. 
Quantitatively, the transition lines are shifted towards higher
fields and lower Zeeman couplings.  In the case $N=6$, 
half of the lowest-energy states even vanish completely when the LLM is taken into account. 

The shift towards higher $B$  means that in the LLM case the states are more
stable against radial expansion as $B$ is increased. ($\Delta M$ increases
as one moves to the right in the figures.)
This should be compared with Ref.\ \onlinecite{bruce2000PRB4718}, where similar
results have been obtained within the LLL approximation, by
assuming the QD to have finite thickness.
Both effects are results of the expansion of the basis set of the Hilbert
space, but in our case the system still remains two-dimensional. 
The vanishing of some lowest-energy states can be explained by noting that 
the two-body correlation factor has strongest effect on the states
that are most compact (\textrm{i.e.}, have the lowest angular momenta). 
For the MDD state, the gain in energy is largest, and it can
completely block some neighboring states whose energy is only
slightly lower in the LLL approximation. 

The value of $\Delta {E_z}$ (corresponding to $g^{\ast}=-0.44$)
for GaAs is $12.7\,\mathrm{\mu eV/T}$. This is a maximum value
(within approximations and assumptions used) that can only be lowered,
using tilted field experiments.
Since the value is much lower than $\Delta E^{\ast}_z$, it should be possible to
encounter partially spin-polarized post-MDD states in experiments,
even without tilting the magnetic field.

In summary, simple trial wave functions for partially and fully
spin-polarized QD systems have been constructed. The wave functions
were shown to produce excellent total energies, and 
provide an accurate estimate of Landau-level mixing.

We demonstrate that the LLM, which is almost always neglected in previous studies, 
is able to suppress the existence of 
certain lowest-energy states with small amount of additional angular
momentum (compared to $N$). Furthermore, a strong shift of transition 
points towards higher magnetic fields due to the LLM is observed. 
Another important result is, 
that despite the LLM, partially spin-polarized states can exist in the post-MDD
region with realistic Zeeman coupling strengths.
The most visible of the these states is likely to have a  single spin flipped and angular
momentum equal to $L_{\text{MDD}} +N-1$.

This research has been supported by the Academy of Finland through its Centers of Excellence Program (2000--2005).

\newcommand{\noopsort}[1]{} \newcommand{\printfirst}[2]{#1}
  \newcommand{\singleletter}[1]{#1} \newcommand{\switchargs}[2]{#2#1}
  \newcommand{\ABSTRACTunitsep}{\;}

\end{document}